\documentclass[prd,aps,a4paper,superscriptaddress,twocolumn,nofootinbib]{revtex4}
\usepackage{graphicx}
\usepackage{color}
\usepackage{dcolumn}
\usepackage{bm}
\usepackage{slashed}
\usepackage{amsmath}
\usepackage{latexsym}
\usepackage{amssymb}
\usepackage{mathrsfs}
\usepackage{amsfonts}
\usepackage{url}
\allowdisplaybreaks

\begin{document}
\title{Ensemble of Deep Convolutional Neural Networks for real-time gravitational wave signal recognition}

\author{CunLiang Ma}
\affiliation{School of Information Engineering, Jiangxi University of Science and Technology, Ganzhou, 341000, China}

\author{Wei Wang}
\affiliation{School of Information Engineering, Jiangxi University of Science and Technology, Ganzhou, 341000, China}

\author{He Wang}
\affiliation{CAS Key Laboratory of Theoretical Physics, Institute of Theoretical Physics, Chinese Academy of Sciences, Beijing 100190, China}

\author{Zhoujian Cao
\footnote{corresponding author}} \email[Zhoujian Cao: ]{zjcao@amt.ac.cn}
\affiliation{Department of Astronomy, Beijing Normal University, Beijing 100875, China}
\affiliation{School of Fundamental Physics and Mathematical Sciences, Hangzhou Institute for Advanced Study, UCAS, Hangzhou 310024, China}

\begin{abstract}
With the rapid development of deep learning technology, more and more researchers apply it to gravitational wave (GW) data analysis. Previous studies focused on a single deep learning model. In this paper we design an ensemble algorithm combining a set of convolutional neural networks (CNN) for GW signal recognition. The whole ensemble model consists of two sub-ensemble models. Each sub-ensemble model is also an ensemble model of deep learning. The two sub-ensemble models treat data of Hanford and Livinston detectors respectively. Proper voting scheme is adopted to combine the two sub-ensemble models to form the whole ensemble model. We apply this ensemble model to all reported GW events in the first observation and second observation runs (O1/O2) by LIGO-VIRGO Scientific Collaboration. We find that the ensemble algorithm can clearly identify all binary black hole merger events except GW170818. We also apply the ensemble model to one month (August 2017) data of O2. There is no false trigger happens although only O1 data are used for training. Our test results indicate that the ensemble learning algorithms can be used in real-time GW data analysis.
\end{abstract}

\maketitle

\section{Introduction}
On September 14, 2015, the advanced Laser Interferometer Gravitational Wave Observatory (aLIGO) directly detected a GW event, GW150914, for the first time in human history \cite{abbott2016observation}. This successful detection directly verified the prediction of the existence of GW in general relativity and opened a new window of universe observation. Such detection opens the era of gravitational wave astronomy. Gravitational wave astronomy is important to the study of the origin and evolution of the universe, the nature of dark matter and dark energy, and others. On August 17, 2017, the gravitational wave event GW170817 \cite{abbott2017gw170817} produced by the merging of a binary neutron star was directly observed. This was the first time that GW, gamma-ray burst, optics and other electromagnetic signals \cite{abbott2017gravitational, goldstein2017ordinary, savchenko2017integral} were clearly and continuously detected from one astrophysical source \cite{2017Multi}. The observations of GW170817 marked the arrival of the era of multi-messenger astronomy including GW.

Gravitational wave astronomy requires multi-disciplinary collaborative research, such as mathematics, physics, and computer science. Since the detected gravitational wave signal is buried in a background noise, elaborate data analysis is needed to extract the GW signal. The widely used data analysis method is matched filtering. With matched filtering, each data segment in the detected strain is matched to a template bank. The maximum value of the output is the matched signal-to-noise ratio (SNR). When it is bigger than a specified threshold, a GW signal candidate is announced. Subsequently, further inspection methods such as $\chi^2$ time-frequency test \cite{allen2005chi}, multi-detector coincidence test \cite{robinson2008geometric}, and others are used to calculate the false alarm rate of the candidate which is used to determine whether it is a true GW event or not. Since the parameters of the potential GW source are not known in advance, the parameter space of the template bank must be widely searched. The matched filtering method is computationally expensive. The low-latency implementation of the matched filtering method cannot be extended to the 9-dimensional signal manifold \cite{huerta2019enabling}. Recently, researches have shown that the method based on the artificial neural network may replace matched filtering to achieve efficient detection.

As early as the 1980s, many researchers were engaged in the study of neural networks \cite{fukushima1982neocognitron,sanger1989optimal,fukushima1983neocognitron,hecht1992theory}. However, the application range of neural networks was very limited at that time, it was only used to deal with small-size picture recognition (such as handwritten digital recognition \cite{lecun1989handwritten}). Because of the impact of the vanishing gradient problem \cite{hochreiter1998vanishing}, the neural network cannot improve the accuracy through deepening the model structure. This problem seriously hindered the development of the neural network. Situation changed in 2012. The AlexNet won the championship in the ImageNet competition based on the deep Convolutional Neural Network (CNN) \cite{krizhevsky2012imagenet}. At present, the CNN have achieved great success in many fields such as image classification \cite{he2016deep,szegedy2016rethinking,simonyan2014very}, image segmentation \cite{long2015fully,chen2017deeplab,he2017mask}, object detection \cite{szegedy2013deep,redmon2016you,ren2015faster}, natural language processing and speech recognition \cite{hinton2012deep,graves2013speech,graves2014towards}.

In 2018, George et al. \cite{george2018deep} and Gabbard et al. \cite{gabbard2018matching} individually concluded that deep learning can achieve the accuracy as matched filtering in gravitational wave data analysis. Compared to matched filtering, the computational efficiency of deep learning is much higher. Deep learning is expected to be applied to real-time detection of transient GW. Recently, the application of the CNN in GW research has been expanded greatly \cite{chan2020detection,baltus2021convolutional,beheshtipour2021deep,ormiston2020noise,chatterjee2019using,wang2020gravitational,xia2021improved,9663260}.

\begin{figure}
\includegraphics[scale=0.13]{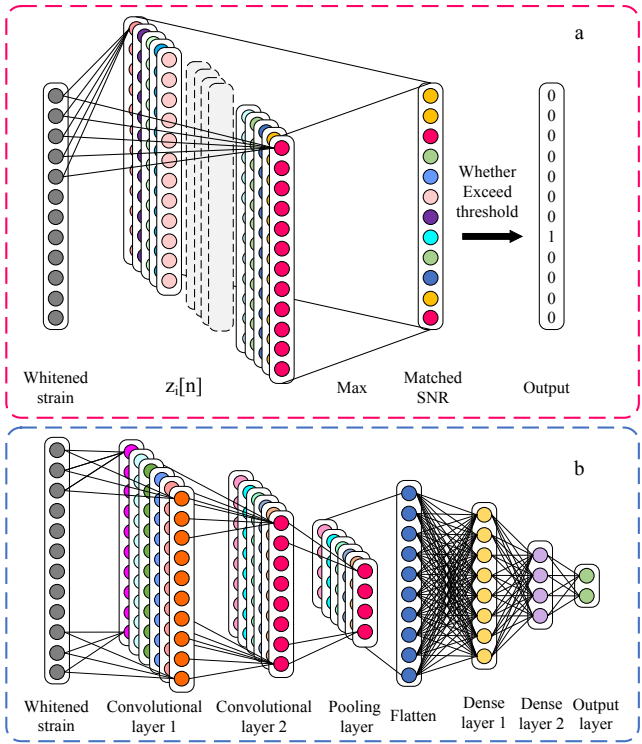}
\caption{\label{fig:1}Diagram of matched filtering (a) and CNN (b). The channel with gray circles represents the whitened strain. In second layer, the various channels with colored circles are the convolutional output of different kernels and the circles in a channel represent the output at different times. Note that the channels of second layer in (a) are not the templates.  In (a), the convolutional kernels of second layer are templates whitened by the background noise while in (b) the kernels are got via training.}
\end{figure}

Fig.~\ref{fig:1} shows the comparison of CNN and matched filtering.  To facilitate the comparative analysis, we give a brief mathematical description of the matched filtering. The matched output corresponding to i-th template in template bank is
\begin{equation}
z_i[n]=T\sum_{m}x[m]h_i[m-n],\label{con:eq1}
\end{equation}
where $T$ is sampling period, $m$ and $n$ represent sampling time, $x[n]$ is the whitened strain, and $h_i[n]$ is the $i$-th template whitened by background noise. Eq (\ref{con:eq1}) indicate that $z_i[n]$ is convolutional results between $x[n]$ and $h_i[-n]$. The matched SNR of the strain can be calculated by
\begin{equation}
z[n]=\max\limits_{i}(z_i[n]),\label{con:eq2}
\end{equation}
From above analysis, we can get a conclusion that matched filtering can be understood as a convolutional neural network with only one convolutional layer composed of a large number of kernels that comprise the template bank and this conclusion is agreed by work of Jingkai Yan et al. \cite{yan2022generalized}. The deep learning model effectively compresses a large number of approximated template waveforms from the dimension of width to the dimension of depth.  Some recent researches show that the deep learning model has a relatively high probability of misjudging the noise as a GW signal \cite{menendez2021searches}. The ensemble of multiple deep learning models may simultaneously focus on the width and depth of the approximated template waveforms and is expected to improve the effect of gravitational wave detection.

Ensemble learning can improve the performance of the learner via a combination of multiple learners. In 1979, Dasarathy et al. \cite{dasarathy1979composite} first proposed the concept of an ensemble system, which opened the prelude to the development of ensemble learning. In 1990, Hansen et al. \cite{hansen1990neural} proved that the ensemble system has the characteristic of variance reduction. The discovery of this property proves that ensemble learning has the ability to improve the effect of neural networks and provides a theoretical basis for neural networks combined with ensemble learning. In 2021, Huerta et al. \cite{wei2021deep} designed a model for GW detection which is composed of two optimized WaveNet. Their model realized real-time signal recognition. Recently, they developed a GW detection workflow with an artificial intelligence (AI) ensemble. Using this AI ensemble, they only spent 7 minutes to analyze all LIGO data in 2017 August. All signals are found without false alarm \cite{huerta2021accelerated}. In the current work, we build an ensemble model to detect real GW events by using the optimized bagging \cite{breiman1996bagging} algorithm and successfully detect almost all GW events in O1 and O2. The most obvious difference between our method and the method in \cite{huerta2021accelerated} is that we individually build two ensemble models for Hanford and Livingston interferometers respectively. Differently, Huerta et al. \cite{huerta2021accelerated} use feature fusion to combine the strain data of two interferometers for each model in the ensemble. The individual model of the ensemble is also called base learner.

The $\chi^2$ time-frequency test together with the matched filtering method can eliminate the false trigger caused by a glitch. However, $\chi^2$ time-frequency test cannot be used together with deep learning. We use the strategy of cross-validation of two detectors to treat this problem. Since the probability of the simultaneous presence of glitch in the detected noise signal from two interferometers is extremely low, combining the detection results from the two interferometers can effectively reduce the number of false triggers caused by the glitch. This is the basic idea for the current work. The test results based on August 2017 LIGO data indicates that our idea works very well.

This paper is organized as follows. In Sec.~\ref{sec2}, we introduce the process of building data for training and testing. After that we describe our ensemble model in Sec.~\ref{sec3}. Then we apply our ensemble model to the O1/O2 LIGO data in Sec.~\ref{sec:4}. The high true alarm rate and extremely low false alarm rate behavior of our ensemble model will reported there. The last section is devoted to a summary.

\section{\label{sec2}data for training and testing}

The dataset for the deep model in this work was generated by the open-source project ggwd \footnote{\url{https://github.com/timothygebhard/ggwd}} \cite{gebhard2019convolutional}. We focus on gravitational wave signals produced by binary black hole (BBH) mergers. Two classes of data are generated. One is the data strain of pure background noise and the other is the data strain including background noise and a GW waveform. The GW waveform contains the complete inspiral, merger, and ringdown phases of the BBH coalescence. The synthetic data can be represented by

\begin{equation}
s(t)=h(t)+ n(t),\label{con:eq3}
\end{equation}
where $h(t)$ represents the GW waveform which is generated by the effective-one-body numerical relativity waveform model (EOBNR) \cite{pan2014inspiral,PhysRevD.95.044028,PhysRevD.96.044028_SEOBNRE,PhysRevD.101.044049_validSEOBNRE}. $n(t)$ is the background noise that is randomly sampled from the O1 data and all of the identified GW events are eliminated.

To filter out the spectral components of the environmental noise and eliminate the influence of Newtonian noise, all the data in the data set were whitened and passed through a high pass filter with a cut-off frequency of 20Hz. The duration of the training data set is 1 second and the sampling rate is 4096Hz.

The mass of both black holes is randomly sampled between $5M_\odot$ and $80M_\odot$, and the dimensionless spin is randomly sampled in (0, 0.998). The polarization angle is sampled uniformly at random from the interval $(0, 2\pi)$. The coalescence phase and the inclination angle are sampled jointly from a uniform distribution over a sphere, while the right ascension and the declination are all sampled from a uniform distribution over the sky. The distance between the Earth and the source is a fixed value of 100Mpc. But the simulated waveforms are later rescaled to match a given network SNR \cite{cutler1994gravitational} which is randomly sampled in (7,20).

A dataset of 120,000 samples was generated, within 45,000 samples containing GW signal and 75,000 samples containing pure background noise. We use 60,000 samples (30,000 GW and 30,000 noise) for training, 10,000 samples (5,000 GW and 5,000 noise) for validation, and 50,000 samples (10,000 GW and 40,000 noise) for testing. In order to mimic the fact that the number of noise samples in actual gravitational wave detection are much larger than that of the gravitational wave samples, we did not adopt the conventional scheme which generate a test set with the 50\% noise samples and 50\% GW samples. Alternatively, we make the test set with 80\% noise samples and 20\% GW samples.

\section{\label{sec3} ensemble deep learning scheme for gravitational wave data analysis}
\begin{figure*}
\includegraphics[scale=0.5]{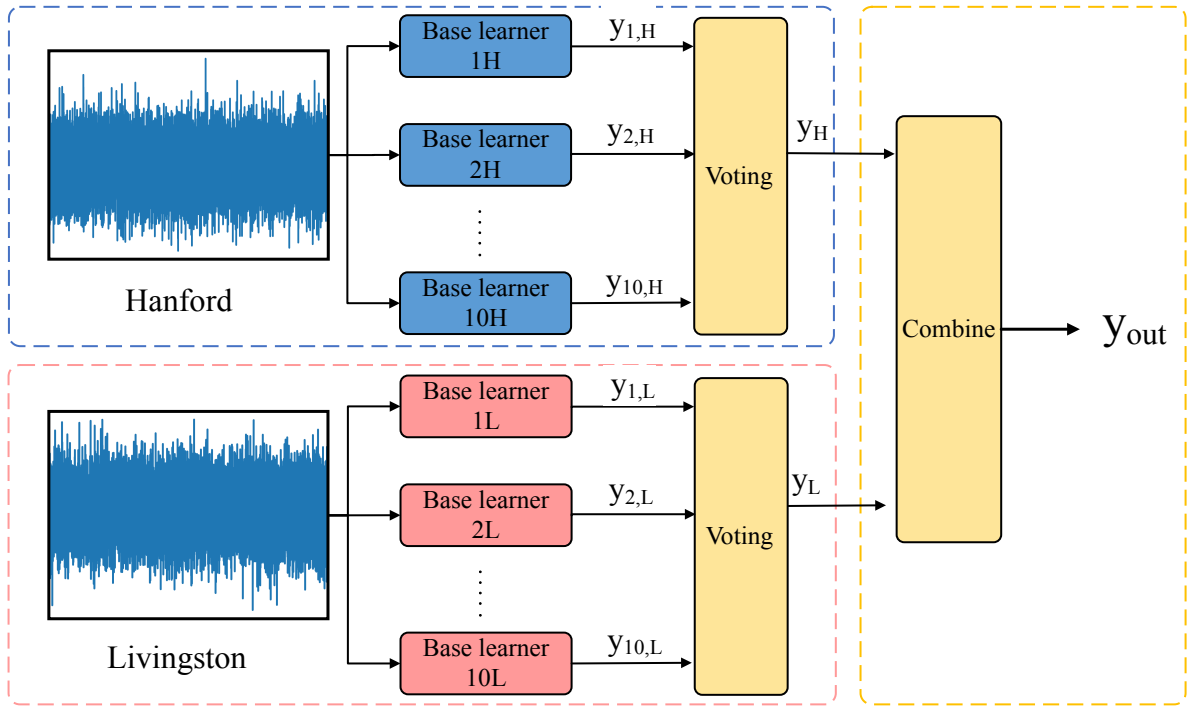}
\caption{\label{fig:2}The structure of the ensemble deep learning model designed in the current work. Base learner $i$H or $i$L ($i\in[1,10]$) denotes the $i$-th base learner for the Hanford or Livingston detector respectively. $y_{i,H}$ and $y_{i,L}$ represents the output of base learner $i$H and $i$L respectively. The output $y_{out}$ of whole model is combined by the outputs $y_{H}$ on Hanford detector and $y_{L}$ on Livingston detector.}
\end{figure*}

\begin{table}[htb]
\caption{\label{tab:table1} The basic structure of base learner.}
\begin{center}
\begin{ruledtabular}
\begin{tabular}{ccc}
Layer&Input&Vector (size: 4096)\\
\hline
1 & Reshape & Matrix (size: $1\times4096$)\\
2 & Convolution & Matrix (size: $ 8\times4033$)\\
3 & Convolution & Matrix (size: $ 8\times4002$)\\
4 & Max pooling & Matrix (size: $ 8\times500$)\\
5 & Convolution & Matrix (size: $ 16\times469$)\\
6 & Convolution & Matrix (size: $ 16\times454$)\\
7 & Max pooling & Matrix (size: $ 16\times75$)\\
8 & Convolution & Matrix (size: $ 32\times60$)\\
9 & Convolution & Matrix (size: $ 32\times45$)\\
10 & Max pooling &Matrix (size: $ 32\times11$)\\
11 & Flatten & Vector (size: $ 352$)\\
12 & Dense layer & Vector (size: $ 64$)\\
13 & Dropout & Vector (size: $ 64$)\\
14 & Dense layer & Vector (size: $ 64$)\\
15 & Dropout & Vector (size: $ 64$)\\
\hline
 & output & Vector (size: $ 2$)\\
 \end{tabular}
 \end{ruledtabular}
 \end{center}
\end{table}

This section exhibits the proposed ensemble model for GW data analysis. We apply the bagging algorithm \cite{breiman1996bagging} to build the ensemble model. Firstly, we introduce the structure of the base learners. Secondly, we present the proposed detection method consisting of two independent ensemble models, and each model detects the data of Hanford and Livingston interferometer, respectively. Then the training method of the base learners and the method for the model selection will be detailed.

\subsection{\label{sec:level2}Base learner}
In recent years, many works investigated the application of the CNN to the GW detection \cite{george2018deep,li2020some,lin2020binary,krastev2021detection,fan2019applying}. The CNN here we used contains one input layer, several hidden layers, and one output layer to form the base learners of the ensemble deep learning model.

Considering the good performance of the model designed by Gabbard et al. \cite{gabbard2018matching}, we adopt this model as the base learner of our ensemble model. In the original work \cite{gabbard2018matching}, the simulated Gaussian noise was used. After using the optimized ensemble method and real background noise, an ensemble model with better detection capability can be constructed. We find that the ensemble model works well for real background noise and surmise that the dropout \cite{srivastava2014dropout} augments the diversity of the base model to make it suitable to be assembled.

The public data recorded by LIGO includes two types which correspond to sampling rate 4096Hz and 16384Hz. In the current work we treat the 4096Hz one. The model designed in \cite{gabbard2018matching} is only applicable to data with the sampling rate 8192Hz, we make some modifications to the structure of their model. Our base learner has six convolutional layers, three max-pooling layers, and two dense layers, which are shown in Table.~\ref{tab:table1}. The nonlinear activation function Elu is used after every layer except the last dense layer. The softmax activate function is added after the last dense layer. The output of the base learner returns a value in the range [0, 1] which indicates whether a BBH coalescence signal is present or not. To further enhance the diversity of the base learners, bootstrap and a data augmentation method are used. We use adaptive moment estimation with incorporated Nesterov momentum \cite{dozat2016incorporating} with a momentum schedule of 0.004. The learning rate is set to 0.002, and the batch number value is 32.

\subsection{\label{sec:level2}Ensemble model}
The architecture of the whole ensemble deep learning model is shown in Fig.~\ref{fig:2}. This ensemble model consists of two individual sub-ensemble models which are used for Hanford and Livingston interferometers respectively. Each sub-ensemble model includes 10 base learners. These 10 base learners are chosen from 100 trained weak learners. The model training and selection methods will be described in the next section.

Here we describe the voting strategy used in the ensemble model. Both hard voting and soft voting are widely used in machine learning. Hard voting uses multiple individual models to make its predictions. Soft voting relies on probabilistic outcome values generated by classification algorithms. Compared to hard voting, soft voting takes into account more information. Based on this consideration our ensemble model adopts the soft voting scheme.

The average voting method can reduce the bias caused by abnormal voting results to a certain extent. In ensemble learning, average voting method is often used. We investigated two types of average voting schemes including arithmetic mean and harmonic mean. The detection capability of the two methods will be shown in the following subsection.

We denote the output of the i-th base learner in the ensemble model belonging to the interferometer $X$ (Hanford or Livingston) as $y_{i,X}$, and the ensemble output as $y_X$. Then for the arithmetic mean, the average can be expressed as
\begin{equation}
y_X=\frac{1}{N}\sum_{i=1}^{N}y_{i,X}.\label{con:eq4}
\end{equation}
For the harmonic mean, the average can be expressed as
\begin{equation}
y_X=\frac{N}{\sum_{i=1}^{N}\frac{1}{y_{i,X}}}.\label{con:eq5}
\end{equation}
In the current work $N=10$ is the number of the base learners of each detector. The detail of the model combination is introduced in the following subsection.

\subsection{\label{sec:level2}Model optimization}
We use several skills to optimize our ensemble model. Some are about the training data. Some are about model construction. And others are about model parameter setting. The following sub-subsections are for these optimizations.

\subsubsection{Augmenting training data}
\begin{figure}
\includegraphics[scale=0.53]{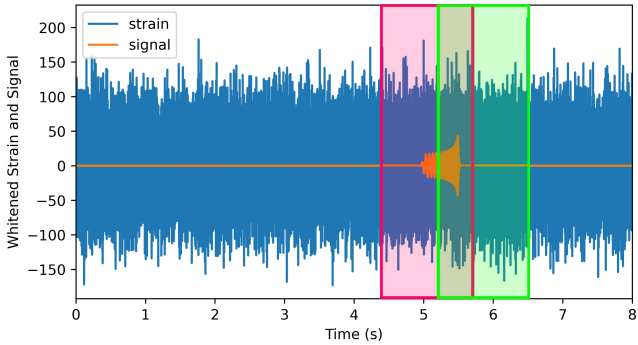}
\caption{\label{fig:3}Extracting one second duration segment from 8 seconds long strain. The red and green blocks represent two examples of extraction. The blue curve is a strain that consists of GW waveform plus noise. The orange curve is the theoretical GW waveform in the strain.}
\end{figure}
\begin{figure*}
\includegraphics[scale=0.5]{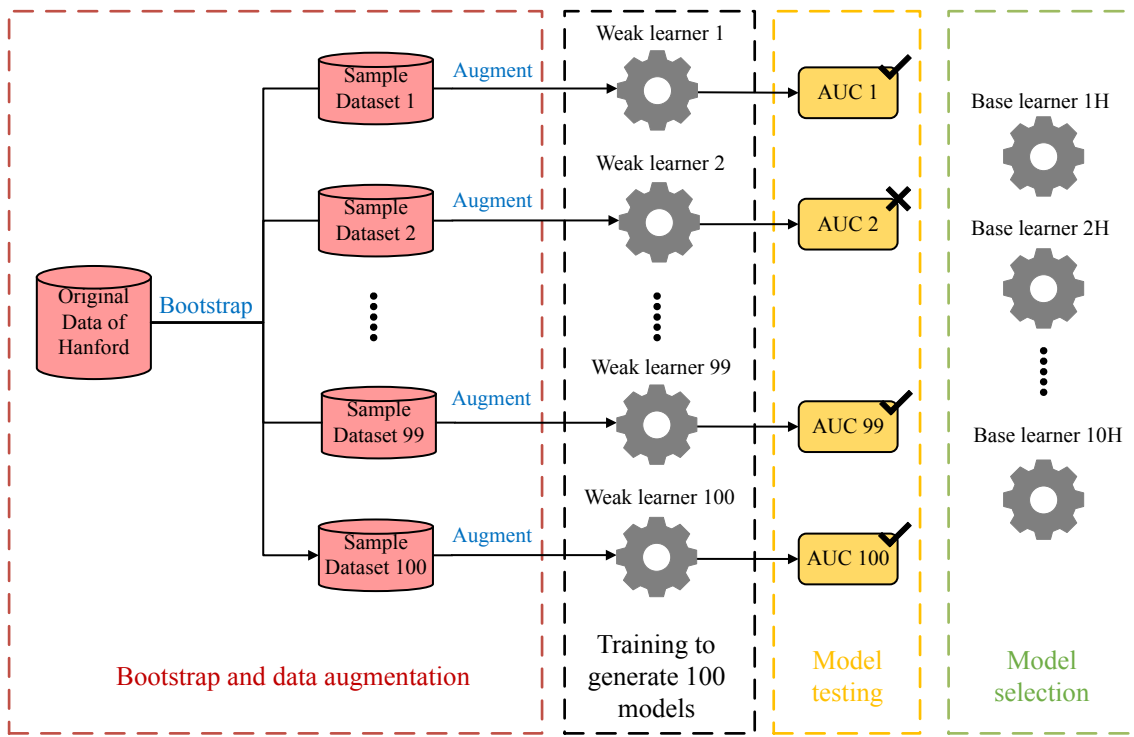}
\caption{\label{fig:4}Schematic diagram of building process of base learners. There are four steps to build the base learners. First step is using bootstrap and data augmentation to build 100 training sets. Second step is training 100 weak learners.  Third step is using validation set to calculate the AUC of each weak learner. Final step is selecting top ten weak learners with largest AUC as base learners.}
\end{figure*}
In order to generate training data, we firstly use ggwd \cite{gebhard2019convolutional} to produce 60,000 samples (30,000 GW and 30,000 noise). Each sample data contain 8 seconds long strain. When we inject a simulated GW signal into the strain, we always let the merger time of the BBH coalescence waveform locate at a fixed time. Specifically the merge occurs at 5.5 s in all GW strains. We named these data as original training set. Similarly, we generate 10000 samples (5000 GW and 5000 noise) for validation and we named these data as original validation set. When we generate the noise for the training and validation data, only O1 data of LIGO are used. Although the training data is quite limited in this noise respect, the performance of our ensemble model is quite good as we will show in the following. This fact also indicates the robustness of our ensemble model for GW data analysis.

For training, We use bootstrap \cite{efron1992bootstrap} to generate 100 sample data sets from the original training set. The detail of the bootstrap operation is as following. Firstly, we divide the 60,000 samples into 600 groups. Each group contains 100 samples. Then we randomly select group from these 600 groups. After each selection, the selected group will be put back and randomly select again. Overall we do 600 selections. After these 600 selections we get a data set which contains 60,000 samples. We repeat the above selection 100 times. Then we get 100 data sets. Each set contains 60,000 samples.

Now we have $100\times60,000$ data samples. Each sample contain 8s long strain. Based on each sample, we randomly extract two segments with duration 1 second and let the merger time of the BBH waveform locate in the range $(\frac{1}{8},\frac{7}{8})$s. In Fig.~\ref{fig:3} we show two examples of such extraction.

For validation, We adopt the same method as above but only randomly extract the data of the original validation set once to generate a validation set containing 10,000 1s long strain.

Through the above described augmentation, we get 100 different training sets (each training set contain 120,000 data samples) for training and one validation set containig 10,000 data samples. Each sample contain 1s-strain. And the merger time of the BBH waveform happens randomly between $\frac{1}{8}$s and $\frac{7}{8}$s.

\subsubsection{Base learner construction}
For each detector including Hanford or Livingston, we train 100 models individually. Since these models only aim to work for the specific training data, the generalization property may not be good. Consequently these models are traditionally called weak learner.

In machine learning, there are many ways to evaluate the performance of a model. For binary classification problems, the receiver operator characteristic (ROC) curves \cite{fawcett2006introduction} can effectively reflect the model's performance. To draw the ROC curve, we need to calculate the true alarm probability (TAP) and false alarm probability (FAP) of the model. We assume that in the test set, the subset containing GW signals is $S_{\rm GW}$, and the subset only containing background noise is $S_{\rm N}$. In the prediction results, the subset of samples predicted to be GW signal is $S_{\rm pGW}$, and the subset of samples predicted to be noise is $S_{\rm pN}$, then TAP and FAP can be calculated according to the following formula
\begin{align}
{\rm TAP}&=\frac{S_{\rm GW}\cap S_{\rm pGW}}{S_{\rm GW}},\label{con:eq6}\\
{\rm FAP}&=\frac{S_{\rm N}\cap S_{\rm pGW}}{S_{\rm N}}.\label{con:eq7}
\end{align}
When the output of the model $p$ is bigger than a given threshold $P_c$, the input sample is classified as GW signal $s\in S_{\rm pGW}$. In this way $S_{\rm pGW}$ is affected by the threshold $P_c$. Along with the changing of $P_c$, corresponding TAP and FAP can be obtained.

The building process of the base learners is shown in Fig.~\ref{fig:4}. For each detector (Hanford or Livingston), 10 base learners are selected from 100 weak learners, and 100 weak learners are trained by the 100 datasets generated in the previous section. In the training process of the weak learner, 10\% of the data in the training set corresponding to the weak learner is randomly selected to verify the weak learner, and we reserve the weak learners with the highest verification accuracy. Note that this part of the validation data does not belong to the validation set used in the next stage, but to the training set corresponding to each weak learner. After getting 100 weak learners, we calculate the area under curve (AUC) of each weak learner based on the validation set. AUC means the area under the ROC curve. AUC has been widely used to reflect the generalization ability of a model \cite{wang2020gravitational}. We select the top ten models with the largest AUC in the 100 weak learners as the base learner.
The code implementation of the above weak learner training and base learner choice is based on the Keras framework \cite{geron2019hands}. All of the computations are done on a NVIDIA 1080Ti GPU. Based on such hardware, the typical delay time of our ensemble model is less than 2 mini-seconds.

\subsubsection{Sub-ensemble model construction}
As explained before and shown in Fig.~\ref{fig:2} our whole ensemble model includes two sub-ensemble models corresponding to Hanford and Livingston detectors. Here we describe how to construct these two sub-ensemble models.

Using the aforementioned method of weak learner training, selecting, and assembling, we get an ensemble model consisting of 10 base learners. In this section, the performance of the single Gabbard model (Model $\uppercase\expandafter{\romannumeral1}$) \cite{gabbard2018matching}, the ensemble model using the arithmetic mean voting strategy (Model $\uppercase\expandafter{\romannumeral2}$) and the ensemble model using the harmonic mean voting strategy (Model $\uppercase\expandafter{\romannumeral3}$) are compared.

\begin{figure}
\leftline{\textbf{(a)}}
\includegraphics[scale=0.55]{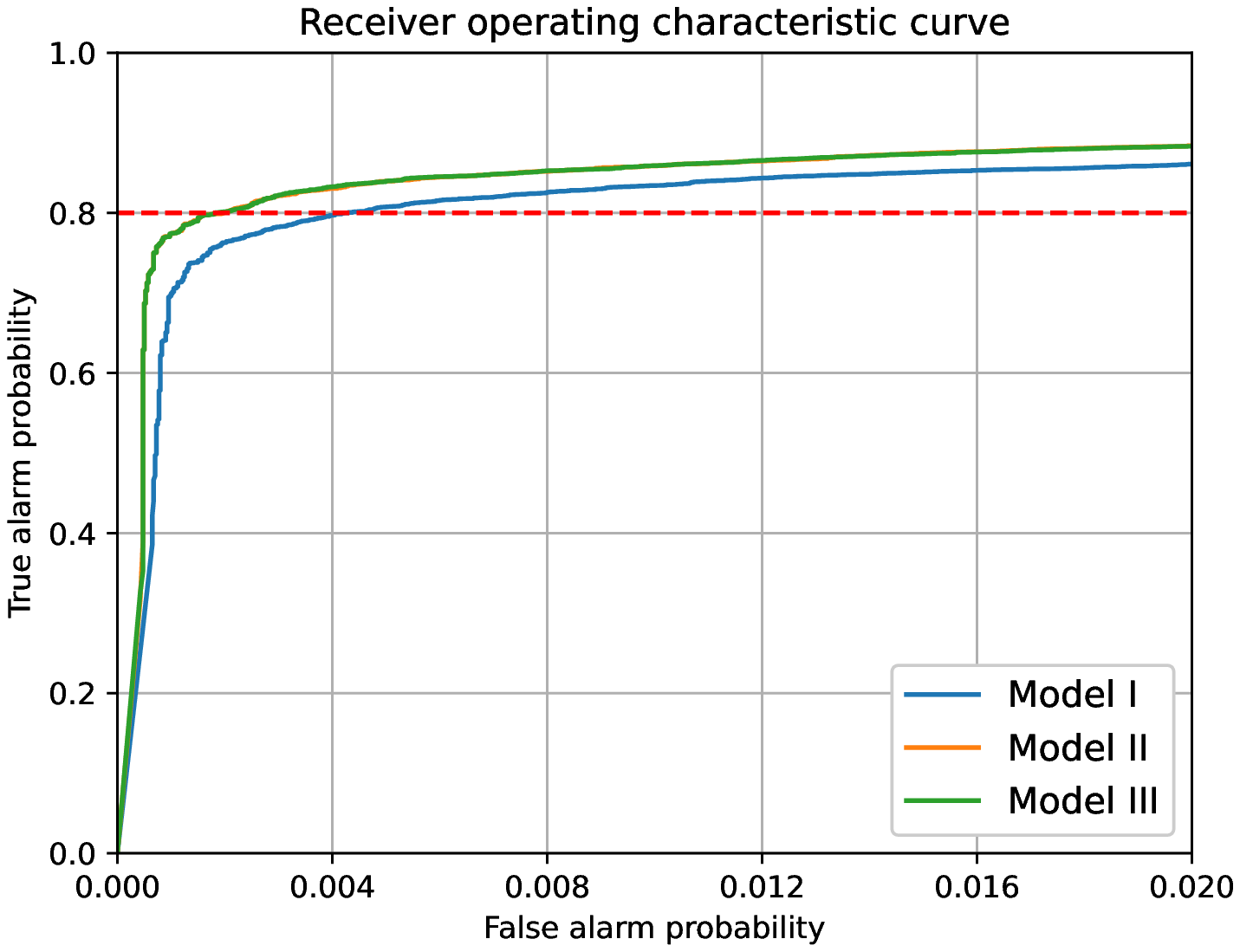}\\
\leftline{\textbf{(b)}}
\includegraphics[scale=0.55]{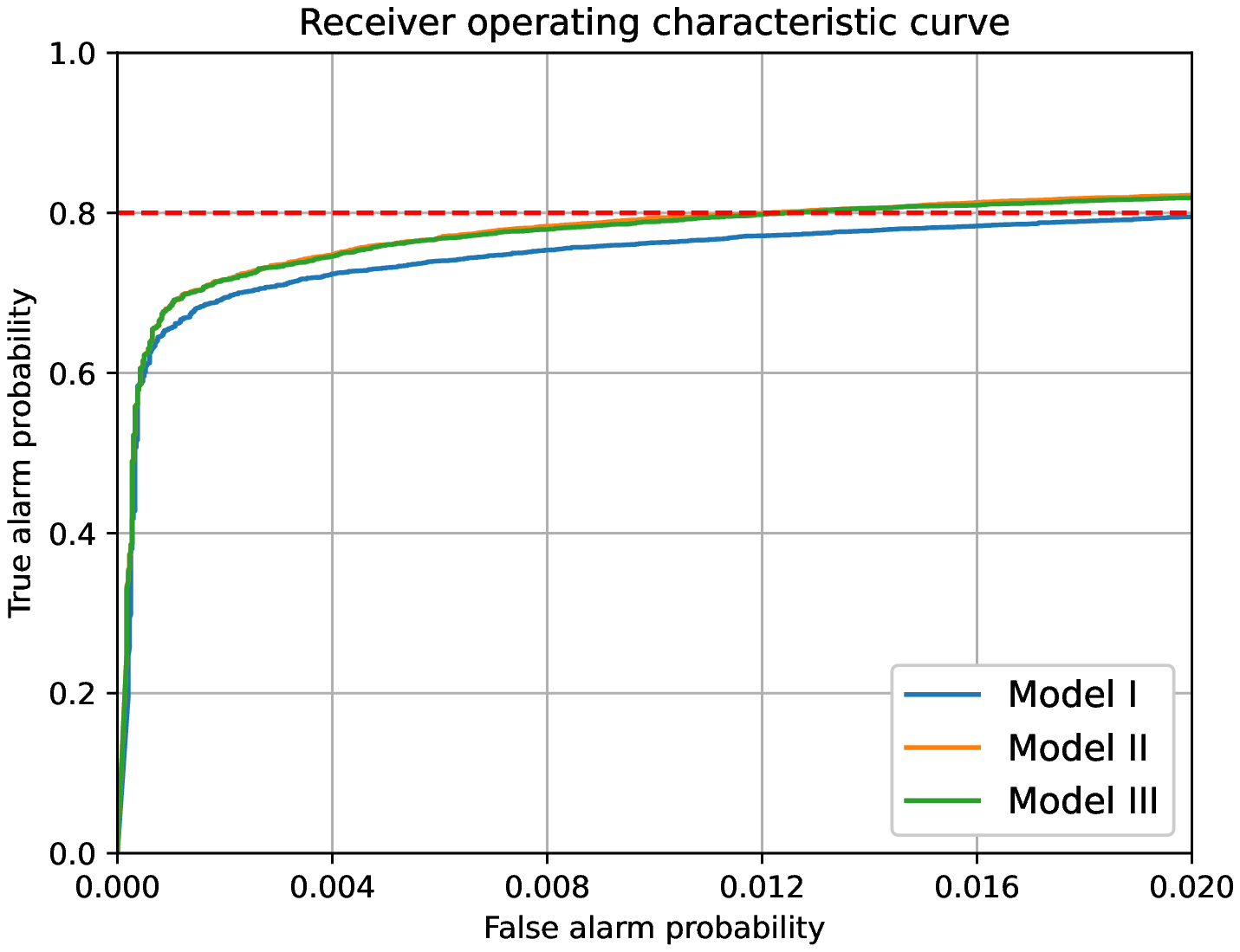}
\caption{\label{fig:5}Receiver operating characteristic (ROC) curve of sub-ensemble models and single model trained by Hanford (a) and Livingston (b) interferometer data. The TAP of red line is 0.8. The blue, orange, and green line indicates the ROC curve of Model $\uppercase\expandafter{\romannumeral1}$, $\uppercase\expandafter{\romannumeral2}$ and $\uppercase\expandafter{\romannumeral3}$ respectively.}
\end{figure}

\begin{figure}
\leftline{\textbf{(a)}}
\includegraphics[scale=0.55]{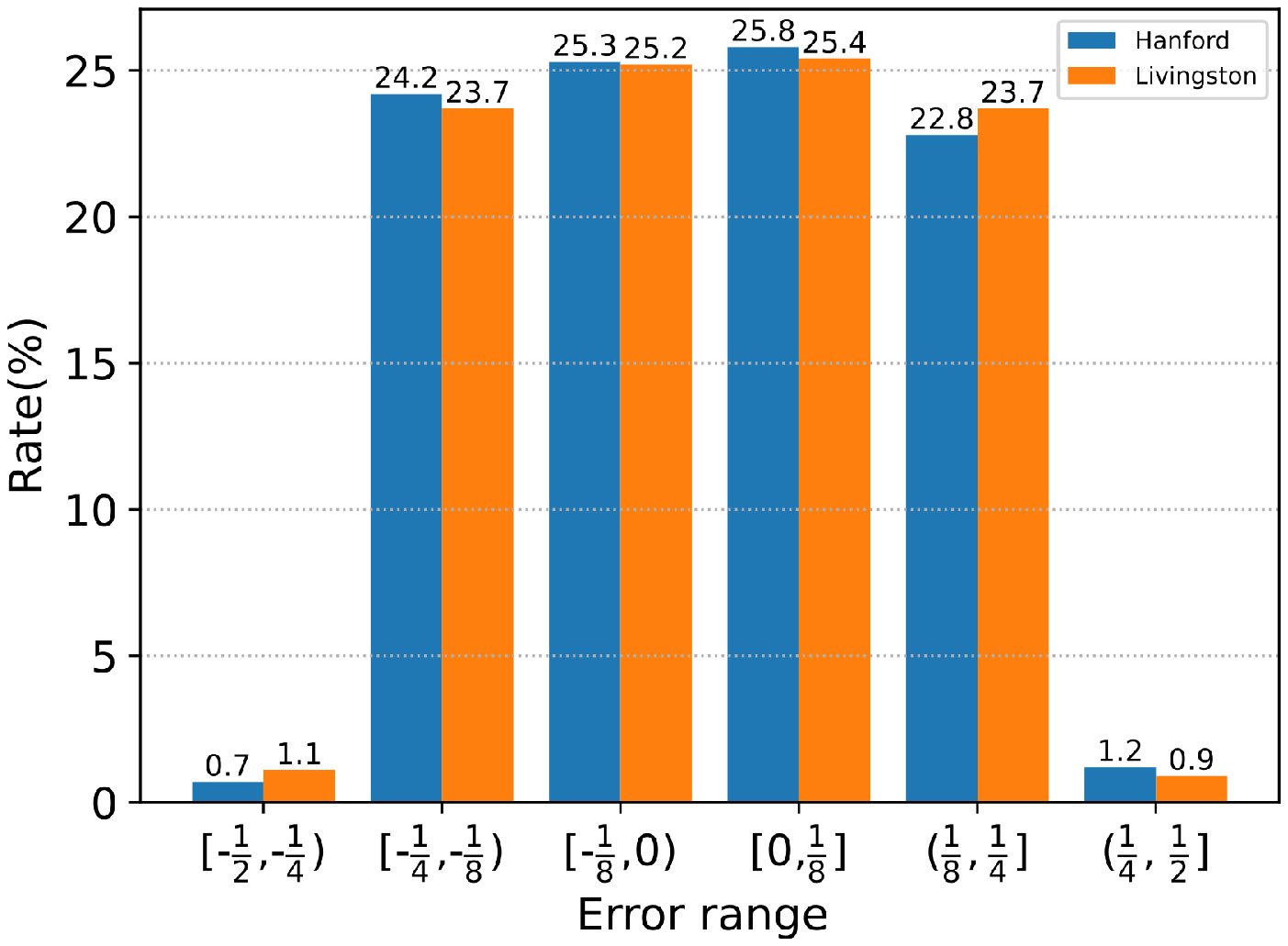}
\\
\leftline{\textbf{(b)}}
\includegraphics[scale=0.55]{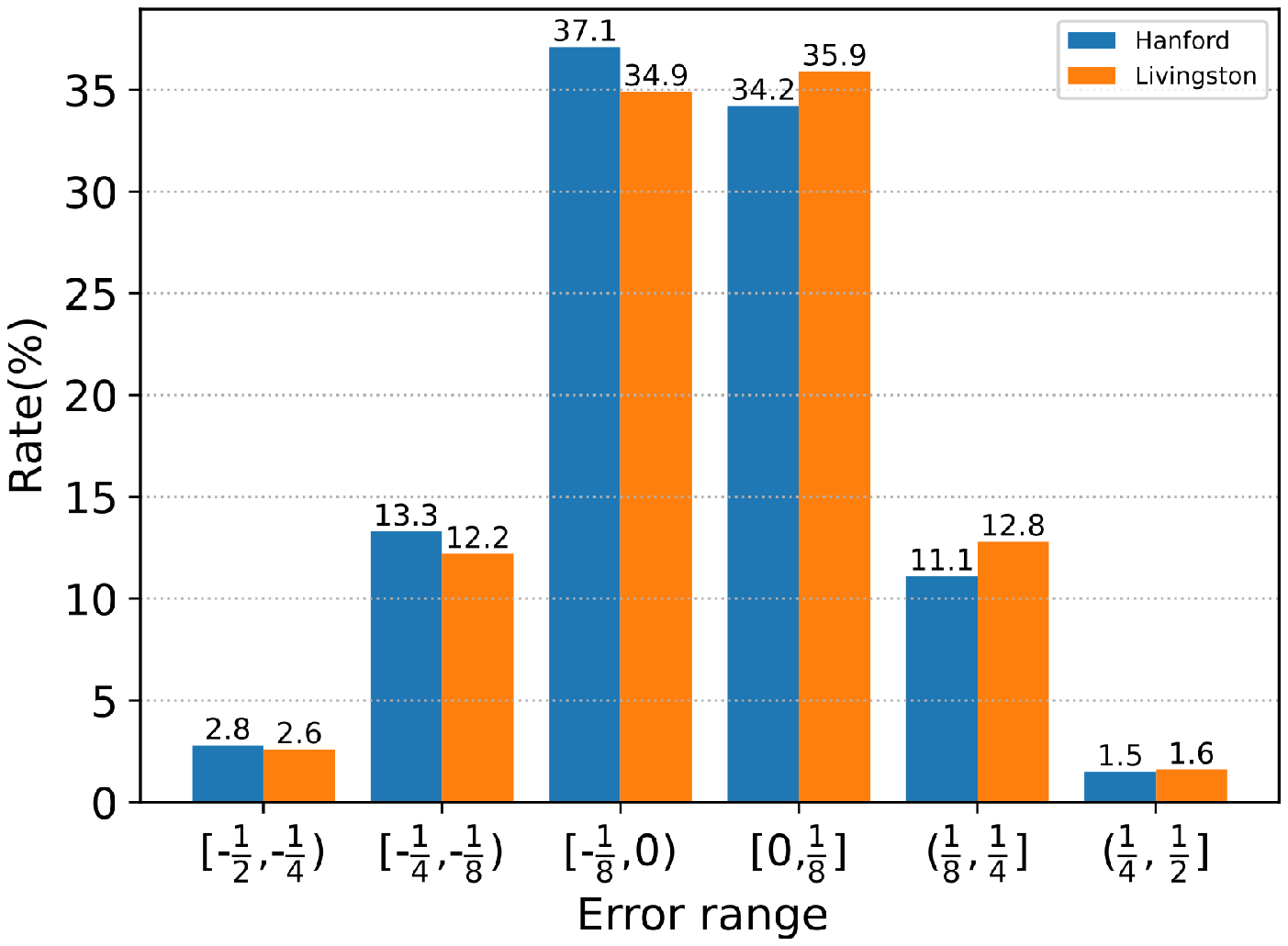}
\caption{\label{fig:6}The distribution histogram of the prediction error of the merger time. Model $\uppercase\expandafter{\romannumeral3}$ is used here. Blue line and orange line are respectively for Hanford and Livingston interferometer. (a) and (b) correspond to moving time step $\Delta t=\frac{1}{8}$s and $\frac{1}{16}$s, respectively.}
\end{figure}
Most of the strain data recorded by LIGO do not include GW signal. Even for current LIGO, the recorded data in most of the time are the background noise without signal. Due to the data characteristics with extremely asymmetrical distribution, the main problem in deep learning for GW data analysis is false triggering. In long duration data, the model may generate numerous false triggers even its false alarm probability is relatively low. The research of deep learning models with low false alarm probability is helpful to develop a gravitational wave detection model with a low number of false alarm triggers. Therefore, we pay much attention to the ROC curves of the low FAP interval.

The ROC curves of Model $\uppercase\expandafter{\romannumeral1}$, Model $\uppercase\expandafter{\romannumeral2}$ and Model $\uppercase\expandafter{\romannumeral3}$ are drawn in the range FAP$\in [0, 0.02]$ which are shown in Fig.~\ref{fig:5}. The ROC curves of Model $\uppercase\expandafter{\romannumeral2}$ and Model $\uppercase\expandafter{\romannumeral3}$ are almost the same in this range. For the results on Hanford interferometer (Fig.~\ref{fig:5}(a)), when the TAP is 0.8, the FAP of the Model $\uppercase\expandafter{\romannumeral1}$, $\uppercase\expandafter{\romannumeral2}$ and $\uppercase\expandafter{\romannumeral3}$ are 0.0042, 0.00195 and 0.00185, respectively, which indicate that with the same TAP, the FAP of the Model $\uppercase\expandafter{\romannumeral2}$ and Model $\uppercase\expandafter{\romannumeral3}$ are about 54\% and 56\% lower than that of the Model $\uppercase\expandafter{\romannumeral1}$. For Livingston interferometer (Fig.~\ref{fig:5}(b)), when the TAP is 0.8, the FAP of the Model $\uppercase\expandafter{\romannumeral1}$, $\uppercase\expandafter{\romannumeral2}$ and $\uppercase\expandafter{\romannumeral3}$ are 0.0025 ,0.0012 and 0.0012 respectively, and in this case the FAP of the Model $\uppercase\expandafter{\romannumeral2}$ and Model $\uppercase\expandafter{\romannumeral3}$ are reduced by about 52\% compared to the Model $\uppercase\expandafter{\romannumeral1}$.

From the above results, we can see that ensemble models have lower FAP than single models with the same TAP. The use of our ensemble model can greatly reduce the number of false triggers in real data analysis. Such improvement can make the GW signal recognition more reliable.

\subsubsection{Moving time step}

For real GW detection data analysis, we face a long-duration data. We take a data segment with duration one second and plug it into our ensemble model for signal recognition. Afterwards we move forward a time step $\Delta t$ to take the following data segment to the following works. The time step will also affect the signal recognition.

Specifically we consider the GW arrival time problem. The gravitational wave arrival time of BBH merger is predicted according to the time range of continuous alarm of GW trigger as
\begin{equation}
t_{\rm arr}=\ \frac{1}{2}\times(t_{\rm fe}+t_{\rm lb}),\label{con:eq8}
\end{equation}
where $t_{\rm fe}$ represents the end time of the detection window of the first alarm, $t_{\rm lb}$ represents the beginning time of the detection window at the last alarm. Since the time position of the GW signal in the training set is symmetrically distributed, we take the midpoint of this interval as the predicted arrival time of the detected gravitational wave. Note that the arrival time (\ref{con:eq8}) does not mean the beginning time of the detected GW signal. Instead, it corresponds to a specific time of the detected GW strain on a given interferometer. Such information is important to be used for determine the location of the GW source.

We use 10,000 samples for each interferometer which contain GW signals in the test set to test the above mentioned arrival time prediction issue. Based on Model $\uppercase\expandafter{\romannumeral3}$ we have tested two moving time steps $\Delta t=\frac{1}{8}$s and $\frac{1}{16}$s. We estimated the prediction error with the difference between the $t_{\rm arr}$ got via Eq.~(\ref{con:eq8}) and the simulated merger time of the binary black hole coalescence waveform. We count the error and plot the histogram in Fig.~\ref{fig:6}. With $\Delta t=\frac{1}{8}$s (Fig.~\ref{fig:6}(a)) and $\frac{1}{16}$s (Fig.~\ref{fig:6}(b)) respectively, Model $\uppercase\expandafter{\romannumeral3}$ successfully detects 8711 and 8751 signals for Hanford interferometer and successfully detects 7851 and 7921 signals for Livingston interferometer.

As we can see in Fig.~\ref{fig:6}(a), among all successfully detected strain in the test on Hanford and Livingston interferometer, 51.1\% and 50.6\%  predictions admit errors falling in $[-\frac{1}{8},\frac{1}{8}]$, and 98.1\% and 98\% predictions admit errors falling in $[-\frac{1}{4},\frac{1}{4}]$. In Fig.~\ref{fig:6}(b), of all successfully detected strain in the test on Hanford and Livingston interferometer, 71.3\% and 70.8\% predictions admit errors falling in $[-\frac{1}{8},\frac{1}{8}]$, 95.7\% and 95.8\% predictions admit errors falling in $[-\frac{1}{4},\frac{1}{4}]$.

Above results indicate that the predicted GW arrival time (\ref{con:eq8}) corresponds to the merger time of BBH. We can understand such correspondence in the following way. BBH merger results in the strongest strain for the whole coalescence process. Consequently the merger time will appear at the center of the continuous triggers produced by our ensemble model. Such fact makes the estimated time (\ref{con:eq8}) corresponds to the merger time of BBH. In addition we find from the above test results that the accuracy of the predicted merger time of BBH GW is improved when the moving time step $\Delta t$ changes from $\frac{1}{8}$s to $\frac{1}{16}$s. We find this is a general trend that smaller moving time step results in more accurate merger time prediction.

\section{\label{sec:4} Application to O1 and O2 data}
\begin{figure*}
\includegraphics[scale=0.43]{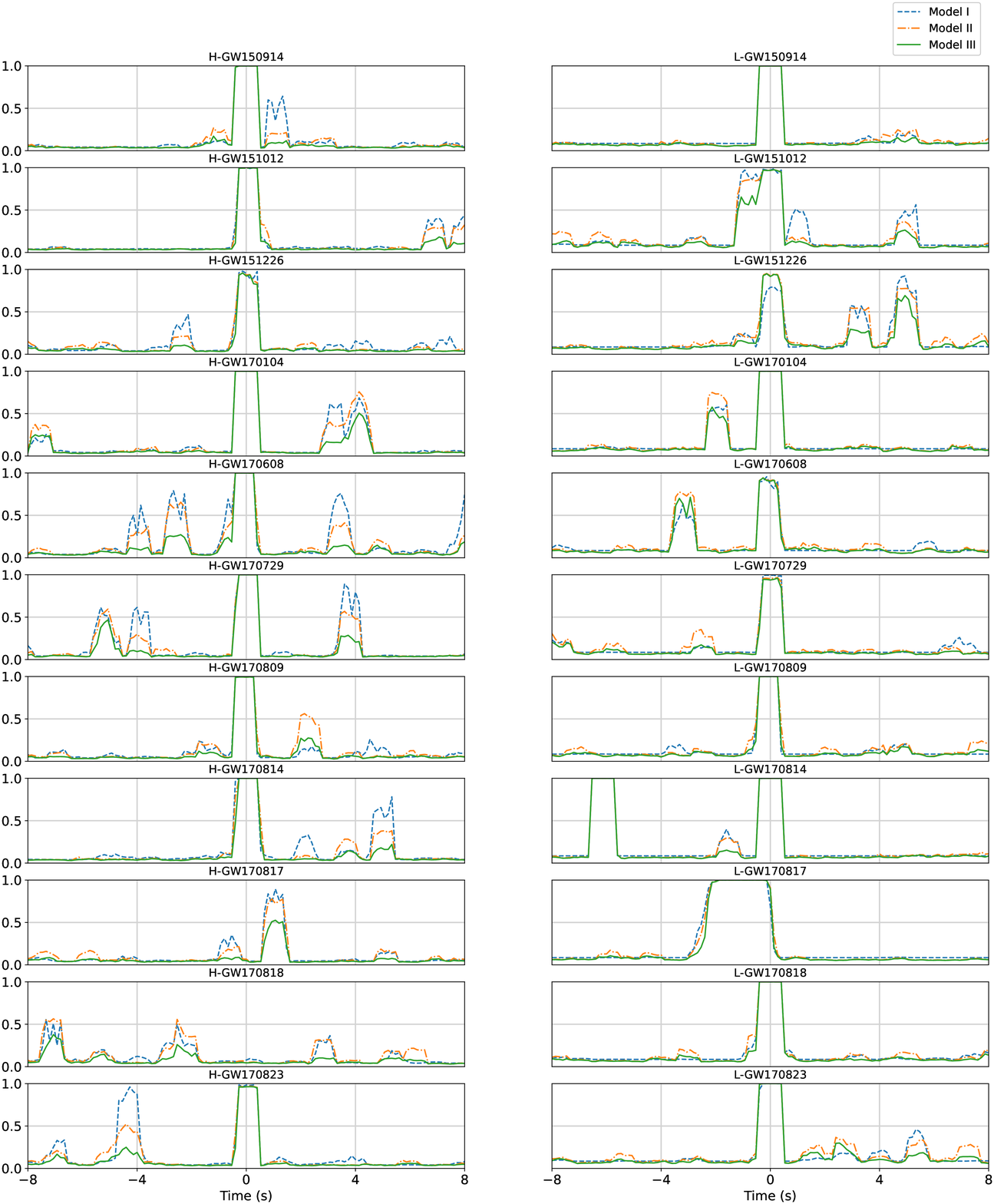}
\caption{\label{fig:7}The GW signal recognition results for the events of O1 and O2. The blue, orange and green line indicates the response of the Model $\uppercase\expandafter{\romannumeral1}$, $\uppercase\expandafter{\romannumeral2}$ and $\uppercase\expandafter{\romannumeral3}$ respectively. H and L represent the tests done for Hanford and Livingston interferometers respectively. All GW events happended at 0 s.}
\end{figure*}

\begin{figure*}
\leftline{\textbf{(a)}}
\includegraphics[scale=0.27]{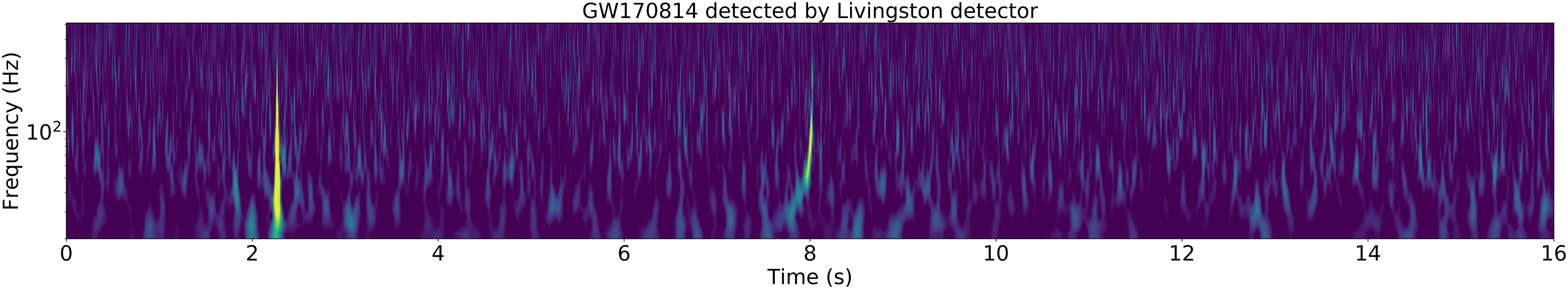}
\leftline{\textbf{(b)}}
\includegraphics[scale=0.27]{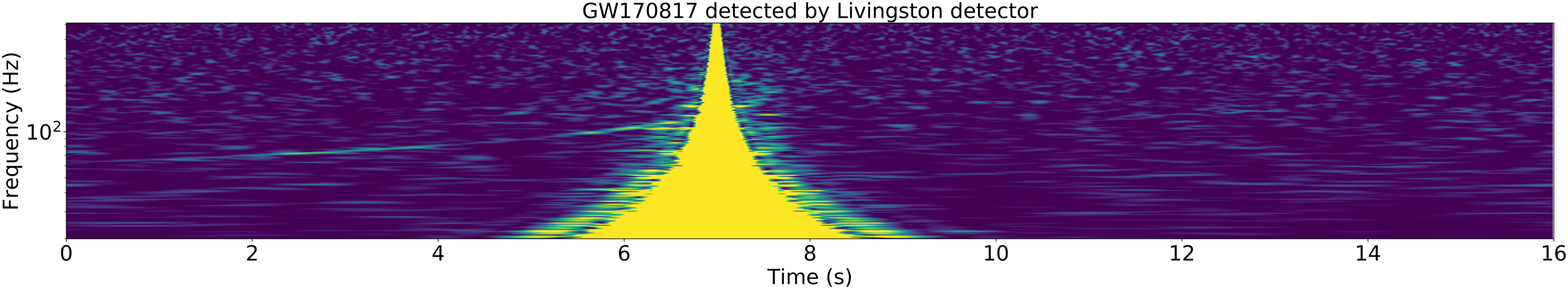}
\caption{\label{fig:8}The spectrogram of 16s strain near L-GW170814 and L-GW170817. Each strain contain a glitch which results in the false trigger of the ensemble models shown in Fig.~\ref{fig:7}. The GW events happended at 8s. Two type of glitch happended near 2 s and 7 s in (a) and (b) respectively.}
\end{figure*}

Now we apply our ensemble model to the O1 and O2 data of LIGO detectors. As we designed in the last section, the length of data segment window is 1s. We take the moving time step $\Delta t=\frac{1}{8}$s.

We have done two investigations. In the first one, we consider all GW events of O1 and O2 reported by LIGO. For each event we take 16s long data and let the GW event locate at the center of the corresponding strain. In the second investigation, we consider 68 minutes long strain and also let the GW event locate at the center of the corresponding strain.

\subsection{Threshold setting}
GW signal recognition is essentially a binary classification problem. Many researchers use a strategy with a detection threshold of 0.5 \cite{wang2020gravitational,xia2021improved}. The background noise related to Hanford and Livingston interferometers has different statistical property. Consequently, the models trained by different detector noise admit different character. Respect to the same threshold, two ensemble models working for Hanford and Livingston interferometer respectively have different FAP and TAP. Accordingly we chose different thresholds for the two ensemble models.

We decide the threshold of the two ensemble models through fixing the FAP to 0.004. Through the ROC curve obtained above, we calculate the thresholds corresponding to model $\uppercase\expandafter{\romannumeral1}$, model $\uppercase\expandafter{\romannumeral2}$ and model $\uppercase\expandafter{\romannumeral3}$ respectively. The thresholds are shown in the table below.  When the output of a sub-ensemble model exceeds its corresponding threshold, an alarm will be given.
\begin{table}[htb]
\caption{\label{tab:table2}The thresholds of the three models on Hanford and Livingston interferometers.}
\begin{center}
\begin{ruledtabular}
\begin{tabular}{ccc}
Detector & Hanford & Livingston\\
\hline
Model $\uppercase\expandafter{\romannumeral1}$ & 0.9761 & 0.9421\\
Model $\uppercase\expandafter{\romannumeral2}$ & 0.8877 & 0.8793\\
Model $\uppercase\expandafter{\romannumeral3}$ & 0.8363 & 0.8454\\
\end{tabular}
\end{ruledtabular}
\end{center}
\end{table}

\subsection{Performance of sub-ensemble models}
We firstly apply the sub-ensemble model to the single interferometer data. When the number of continuous alarms exceeds 5, we let the model generates a GW trigger.

Regarding to the 16s-strain tests we plot the analysis results in Fig.~\ref{fig:7}. H denotes the strain detected by Hanford and L denotes the strain detected by Livingston. The comparison among Model $\uppercase\expandafter{\romannumeral1}$, $\uppercase\expandafter{\romannumeral2}$ and $\uppercase\expandafter{\romannumeral3}$ is shown there. Except H-GW170818, all BBH GW events in O1 and O2 are successfully recognized by ensemble Model $\uppercase\expandafter{\romannumeral2}$ and Model $\uppercase\expandafter{\romannumeral3}$. As we mentioned before, although the ensemble models are trained based on only O1 data, they work very well for O2 data. In contrast, the single Model $\uppercase\expandafter{\romannumeral1}$ misses H-GW151226, L-GW151226, and L-GW170608 signals. Since the training data do not contain binary neutron star (BNS) coalescence, the BNS gravitational wave event GW170817 is lost. As we can see from the recognition results of H-GW170817, all three types of models produced outputs greater than 0.5 at 1s, but the duration of the outputs did not cover the time of the event and the outputs did not exceed the corresponding thresholds. Similarly, for L-GW170817, although the three types of models have successfully generated GW triggers, the duration of the trigger is more matched with the occurrence time of the glitch shown in Fig.~\ref{fig:8}(b). We think this trigger may be a false trigger.

Interestingly, all models result in a false trigger about 6 s before the event for L-GW170814. We check the constant-Q Transformation \cite{brown1991calculation} of L-GW170814 in Fig.~\ref{fig:8}(a). From this figure, we can see a glitch corresponding to the false trigger near L-GW170814.

Similarly, all models result in a false trigger with an alarm time more than 2s at approximately 4 seconds before the GW event L-GW170817. We again check the constant-Q Transformation of L-GW170817 in Fig.~\ref{fig:8}(b). We can see the famous glitch of GW170817 which result in the false trigger.

Regarding to H-GW170818, none of the models successfully recognize this signal. We remind that the matching SNR of GW170818 in Hanford is only 4.1. GW170818 was only detected by GstLAL \cite{sachdev2019gstlal,messick2017analysis}. PyCBC \cite{usman2016pycbc} missed this signal because the SNRs in Hanford and Virgo are 4.1 and 4.2, respectively, which was lower than the threshold 5.5 \cite{abbott2019gwtc} used by PyCBC. Similarly the training set used in the current work admit lowest SNR 7. Consequently our model also miss this signal.

We have seen above that the sub-ensemble models perform better than single model. In the next sub-section we will see that our whole ensemble model can even improve much more.
\subsection{Performance of the whole ensemble model}
\begin{figure*}
\includegraphics[scale=0.47]{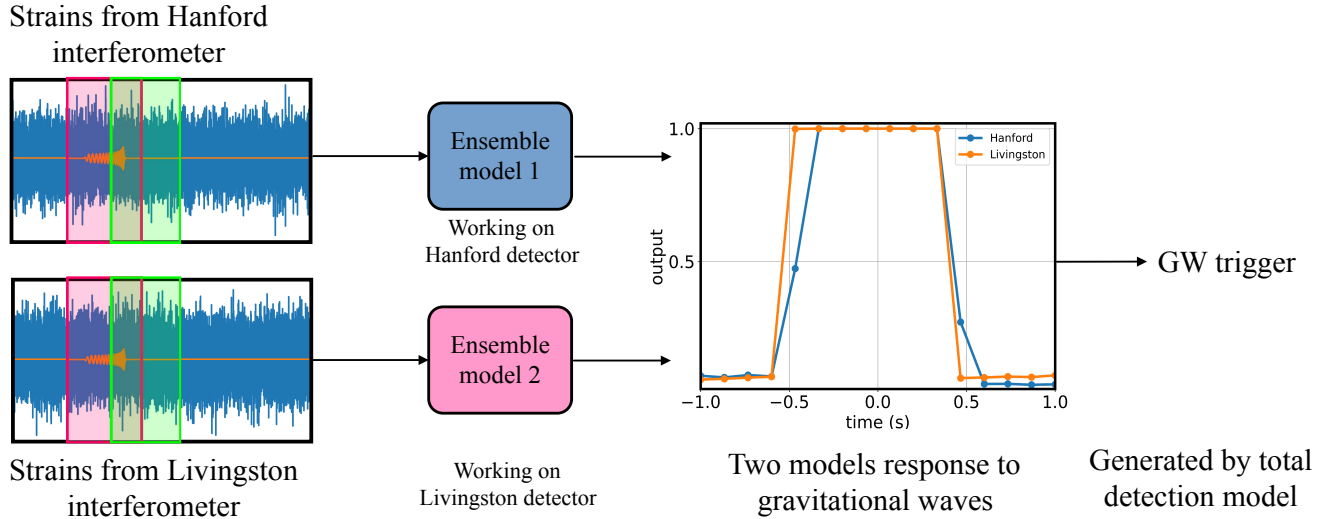}
\caption{\label{fig:9}The schematic diagram of the ensemble model's work style. When the two sub-ensemble models generate more than 5 alerts continuously, a GW signal trigger is given by the ensemble model.}
\end{figure*}

\begin{table}[htb]
\caption{\label{tab:table3}The merge time prediction.}
\begin{center}
\begin{ruledtabular}
\begin{tabular}{ccc}
Event & Predict merge time & Merge time \cite{RICHABBOTT2021100658}\\
\hline
GW170809 & 1186302519.75 & 1186302519.8\\
GW170814 & 1186741861.5625 & 1186741861.5\\
GW170823 & 1187529256.5625 & 1187529256.5\\
\end{tabular}
\end{ruledtabular}
\end{center}
\end{table}

The test results shown in the last sub-section show that the sub-ensemble models corresponding to a single interferometer behaves better than a single model. Now we combine the two sub-ensemble models for Hanford and Livingston interferometers to obtain more reliable real-time GW signal recognition results. In this sub-section, we even increase the difficulty of the problem. We consider 68 minutes long data which is much longer than 16s considered in the last sub-section. The schematic diagram of our ensemble model's work style is shown in Fig.~\ref{fig:9}. If the two sub-ensemble models generate more than 5 alerts continuously, a GW signal trigger will be given. All the tests done in this sub-section are based on Model $\uppercase\expandafter{\romannumeral3}$.

Regarding to false alarm problem, the two individual sub-ensemble models produce averagely 12.85 and 11.07 triggers per hour respectively for Hanford and Livingston interferometers. In contrast, our ensemble model does not give triggers at all except the 10 BBH events. Regarding to the missing trigger problem, our ensemble model fail to find out GW170818. As we analized in the last sub-section, GW170818 is only detected by Livingston interferometer. Consequently our ensemble model misses this signal.

Besides the 68 minutes data segment near the 10 BBH events, we have tested the whole LIGO data in August 2017. Note that we only analyse the time when both Hanford and Livingston interferometers were normally working. After combining the two sub-models, our whole ensemble model does not give any more triggers except the GW events reported by LIGO.

The results of the FAP do not directly correspond to the false-alarm-rates (FAR) on continuous detection via moving the time window. In the continuous analysis of background noise, the FAR reflects the average time for the model generating a false trigger. The FAR is given by \cite{PhysRevD.105.043002}
\begin{equation}
{\rm FAR}=\frac{N_f}{t_0},
\label{con:eq9}
\end{equation}
where $N_f$ represents the number of false triggers generated in detection process and $t_0$ represents the detection duration. However there was no false trigger in the one-month detection by using the whole detection model, we can't estimate the FAR of whole ensemble model directly. Alternatively, we can estimate the FAR from the detection results of two sub-ensemble models. From the Eq.~(\ref{con:eq9}), we can get
\begin{equation}
{\rm FAR}=\frac{N_f/N}{t_0/N}=\frac{p_{ft}}{d_s}, \label{con:eq10}
\end{equation}
where $N$ represents the number of detections in the duration $t_0$, $p_{f}=\frac{N_f}{N}$ corresponds to the probability of generating a false trigger and $d_s=\frac{t_0}{N}$ corresponds to the detection time step. Note that $p_{f}$ is not FAP. FAP only considers the output of a single detection rather than GW trigger (more than 5 continuous alerts). Assuming that Hanford and Livingston interferometers have independent distributions of background noise data, we can conclude that
\begin{equation}
p_{f}=p_{fH}\times p_{fL},
\label{con:eq11}
\end{equation}
where $p_{fH}$ and $p_{fL}$ represent the probability of generating a false trigger by the sub-ensemble model working on Hanford and Livingston interferometers respectively. Based the Augest 2017 detection result of two sub-ensemble models, we find that $p_{fH}\approx4.16\times{10}^{-4}$ and $p_{fL}\approx3.09 \times{10}^{-4}$. Through Eq.~(\ref{con:eq11}), $p_{f}\approx1.29 \times{10}^{-7}$. Consequently the whole ensemble model's FAR is about 1 per 11.24 days. The coworking time of Hanford and Livingston interferometers in August 2017 is about 17 days which means the expected false alarm event is about 1. That is consistent to the fact that there is no false trigger happening at all.

We also estimate the arrival time of the three BBH merger events by Eq.~(\ref{con:eq8}) and list the result in Table.~\ref{tab:table3}. We find that the prediction error of the three GW events is less than 0.1 s.
This means our ensemble model can roughly recover the data analysis results based on complicated Bayesian method. Based on these good performances, we conclude that our ensemble model can be well applied to real time GW signal recognition.

\section{\label{sec:level1}SUMMARY}
In this work, we designed an ensemble model for gravitational wave signal recognition. The whole ensemble model consists of two individual sub-ensemble models. The model is trained by O1 data of Hanford and Livingston interferometers, respectively. We applied this ensemble model to the gravitational wave events of O1 and O2 of LIGO. Our ensemble model can successfully figure out 9 BBH GW events of the 10 reported ones by LIGO. The missed GW170818 admits subthreshold behavior for Hanford detector. This means our ensemble model admits very high true alarm rate.

We have also applied our ensemble model to the whole month LIGO data in August 2017. Except the BBH events reported by LIGO, our ensemble model does not give extra GW trigger. This means our ensemble model admits extremely low false alarm rates. These test results indicate that the machine learning with ensemble algorithm \cite{huerta2021accelerated} can greatly reduce the number of false triggers in the GW signal recognition process.

The above mentioned high true alarm rate and extremely low false alarm rate prove that the optimized ensemble learning algorithm can be applied to real-time gravitational wave signal recognition. Compared with feature fusion strategy, cross-validation used in the ensemble model does not require additional training and consequently less computation cost. And cross-validation is more explicable than feature fusion.

As caveats, there are also drawbacks to cross-validation. Some GW signal may be confidently detected by just one detector while the another detector admit low SNR. Such kind of signal will be missed like GW170818.

\section*{Acknowledgments}
%|--------------------------------------------------------------------|
This research has made use of data and web tools obtained from the gravitational-wave Open Science Center, a service of LIGO Laboratory, the LIGO Scientific Collaboration, and the Virgo Collaboration.

This work was supported in part by the National Key Research and Development Program of China Grant No. 2021YFC2203001 and in part by the NSFC (No.~11920101003 and No.~12021003). Z. Cao was supported by CAS Project for Young Scientists in Basic Research YSBR-006.
%|--------------------------------------------------------------------|

\bibliography{refs}

\end{document}